\documentstyle[aps,twocolumn,epsf]{revtex}  
\begin{document}
\twocolumn[\hsize\textwidth\columnwidth\hsize\csname 
@twocolumnfalse\endcsname                            
\title{Effect of anharmonicities in the critical number of trapped
condensed atoms with attractive two-body interaction}
\author{Victo S. Filho$^{1}$, Arnaldo Gammal$^{1,2}$ and Lauro 
Tomio$^{1}$}
\address{$^{1}$Instituto de F\'{\i}sica Te\'{o}rica,
Universidade Estadual Paulista, 01405-900, S\~{a}o Paulo, Brazil \\
$^{2}$Instituto de F\'\i sica, Universidade de S\~ao Paulo,
05315-970, S\~ao Paulo, Brazil}
\maketitle
\begin{abstract}
{We determine the quantitative effect, in the maximum number of 
particles and other static observables, due to small anharmonic terms added 
to the confining potential of an atomic condensed system with negative 
two-body interaction. 
As an example of how a cubic or quartic anharmonic term can affect the
maximum number of particles, we consider the trap parameters and the 
results given by Roberts {\it et al.} [Phys. Rev. Lett. {\bf 86}, 4211 
(2001)]. However, this study can be easily transferred to other 
trap geometries to estimate anharmonic effects.
}\newline{PACS 03.75.Fi, 32.80.Pj} 
\end{abstract}
\vskip 0.5cm ]                              

\section{Introduction}

The experimental realization of Bose-Einstein condensation 
(BEC) in magnetically trapped weakly interacting 
atoms~\cite{and95,brad97,mew95,hidr} brought a lot of interest in 
theoretical studies of the properties of condensed atomic systems.
For systems with negative two-body scattering length, recently it was 
reported in Ref.\cite{JILA} a discrepancy between experimental 
results and theoretical predictions~\cite{varios}, in the 
maximum critical number of trapped atoms, $N_c$.
In Ref.\cite{brief}, it was shown that part of the discrepancy 
can be explained by taking into consideration the axially deformed 
shape of the trap. 
The theoretical prediction of the critical number of 
atoms, for the cylindrical symmetry considered in Ref.\cite{JILA}, has to 
be adjusted to a number that is lower than the number given by the 
corresponding spherically symmetric trap. 
Still, this correction is not enough to obtain a result that 
is totally compatible with experimental values. 
So, it is relevant to look for other possible sources of the observed 
discrepancy, in this specific case; as for example, higher order 
non-linearities in the mean-field approximation~\cite{Gammal}, or 
possible experimental deviations not already taken into account.

The parameter associated with the critical number $N_c$, given in 
Ref.\cite{JILA}, is defined by
\begin{equation} 
k=\frac{N_c\mid a \mid}{\sqrt{\hbar / (m\omega)}},
\label{1}
\end{equation}
where $\omega\equiv(\omega_x\omega_y\omega_z)^{1/3}$ is the 
geometrical mean value of the trap frequencies, $m$ the 
mass of the atomic species, and $a$ the
two-body scattering length of the particles in the condensate.  
With the assumption of a spherical symmetrical trap, 
$k\cong 0.575$~\cite{varios}. 
The critical number of atoms for Bose-Einstein condensates with $a<0$  
have been investigated by the JILA group, taking into account experiments  
with $^{85}$Rb~\cite{JILA}, considering a wide tunning of the scattering 
length $a$, from positive to negative values, by means of Feshbach resonance  
\cite{avaria}. The experimental result for the number $k$, 
$k_{expt} = 0.459\pm$ 0.012 (statistical) $\pm 0.054$ (systematic),  
corresponds to a deviation of about 20\% lower than the predicted spherical 
symmetrical result. Part of this discrepancy was shown to be related to the 
cylindrical cigar-shaped symmetry employed in the JILA's experiment,
where the frequencies corresponding to the directions $z$ and
$r_\perp\equiv\sqrt{x^2+y^2}$ are, respectively, given by 
$\omega_z = 2\pi \times 6.8\; \rm{Hz}$ and 
$\omega_{\perp} = 2\pi \times 17.35\; \rm{Hz}$. So, the parameter of 
anisotropy is given by $\lambda\equiv\omega_z / 
\omega_{\perp}=0.3919$. 
The theoretical prediction, obtained in Ref. \cite{brief}, with 
the correct cylindrical symmetry given in Ref.~\cite{JILA}, is 
$k\cong 0.55$. In this case, the experimental result is smaller 
than the predicted value by about 16.5\%. Such a result is still 
not enough to include the theoretical result within the error 
bars of the experimental result.

As explained in Ref. \cite{Roberts}, in the JILA's experiments 
with $^{85}$Rb, the trap is practically harmonic in the central 
region, for very low temperatures and for small size 
condensates. However, one should check how a deviation of the
harmonic trap, outside of the central region, can affect the 
number of condensed atoms. In this case, we are considering
the possible existence of imprecisions in the form of the confining
potential, generated by the modified Ioffe-Pritchard
design used (known as baseball trap). 
The modification of the ground state solution of 
the condensed state can alter, correspondingly, the observables 
associated with it. 

In our present investigation we consider a deviation in the harmonic trap
potential that is effective only outside the central part of the
potential, with the addition of a term that is proportional to 
a cubic or quartic power of the distance. This work was first 
motivated by looking for a possible source of the observed discrepancy 
between theory and the JILA's experiment. However,
we realize that, if exist any deviation from the
harmonic trap in the experiments reported in Ref.~\cite{JILA}, 
it should be of a very small factor considering the kind of trap 
design used by them~\cite{Roberts}.
But, one should be aware also of other kind of trap arrangements in the 
experimental studies of BEC, where an investigation of possible effects 
in the observables due to anharmonic terms in the interaction can be 
useful. In this perspective, our present study of the effect of anharmonic 
terms added to a harmonic trap interaction, is not restricted to the
example that we are going to consider. We consider the trap parameters 
of the JILA's experiments~\cite{JILA} as an example, estimating 
deviations in $N_c$ due to anharmonicities, that can easily 
be extended to other geometrical trap configurations, with the help of
previous studies~\cite{brief}. Presently, we are reporting numerical 
results obtained for $N_c$ when the trap 
deviates from the harmonic shape. We should also note that there exists a 
previous study considering the occurrence of anharmonic terms in a {\it 
time orbiting potential} trap~\cite{minogin}. 

As shown by our results, a steeper confinement will result in a lower
$k$, against the naive expectation. 
At first sight, a steeper confinement should increase the 
kinetic and potential energies, with stabilization for a larger
(and not smaller) value of $k$. However, a steeper confinement also
means that the relative distance between the atoms is less than in
the case of the harmonic potential. Therefore, with the atoms 
feeling more attraction from the interaction term, this will result
in collapse for smaller $k$.

Next, we review the formalism used in the present approach, 
followed by the main results and conclusions.

\section{Mean field approximation}

The Gross-Pitaevskii equation (GPE) that describes the 
wave-function of the condensate $\Psi$ in the mean-field 
approximation has the form 
\begin{eqnarray}
i\hbar \frac{\partial\Psi}{\partial t}=\left(-\frac{\hbar^{2}}{2m}
\vec{\nabla}^{2}+U_{trap}-\frac{4\pi\hbar^{2}|a|}{m} 
\mid\Psi\mid^{2}\right)\Psi\,,
\label{egp}
\end{eqnarray}
where the potential $U_{trap}\equiv 
U_{trap}(\vec{r};\omega_x,\omega_y,\omega_z)$ is given by a 
modified harmonic oscillator trap, and the wave function 
$\Psi\equiv\Psi(\vec{r},t)$ is 
normalized to the number of atoms $N$.
For the stationary solution of Eq. (\ref{egp}), with 
$\Psi(\vec{r},t)=\exp{\left(-{\rm i}\mu t/{\hbar}\right)}
\Psi(\vec{r},0)$,
where $\mu$ is the chemical potential, we obtain
\begin{eqnarray}
i\hbar \frac{\partial\Psi}{\partial t}=\mu \Psi\,.
\label{stat}
\end{eqnarray}
Considering the symmetry used in the JILA's experiment, 
where the trap frequencies are given by
$\omega_\perp\equiv\omega_x=\omega_y\ne\omega_z$, it is 
adequate to work in cylindrical coordinates: 
\begin{equation}
{r}_{\perp} = \sqrt{x^2+y^2}\;\;\;{\rm and}\;\;\; 
\theta={\rm arctan}\left(\frac{y}{x}\right).
\end{equation}
For the ground state of the condensate, we have
$\Psi\equiv\Psi(r_\perp,z,t)$.
By using the trapping geometrical average frequency
$\omega\equiv\left(\omega_x\omega_y\omega_z\right)^{1/3}$ and 
the oscillator length by $l_o\equiv\sqrt{\hbar/(m\omega)}$, we 
obtain the following dimensionless coordinates and parameters: 
\begin{eqnarray}
\rho&\equiv&\sqrt{2}\frac{r_\perp}{l_o},\;\;\; 
\zeta\equiv\sqrt{2}\frac{z}{l_o},\;\;\;\tau\equiv\omega t;
\nonumber\\
\omega_{\rho}&\equiv&\frac{\omega_\perp}{\omega},\;\;\; 
\omega_{\zeta}\equiv\frac{\omega_z}{\omega}. 
\end{eqnarray}
With the above dimensionless units, redefining $\Psi$, 
\begin{equation} 
\Phi(\rho,\zeta,\tau)\equiv
\sqrt{\frac{4\pi\hbar|a|}{m\omega}}\;
\Psi(\vec{r},t),
\end{equation} 
and the trap potential,
$U_{trap}\equiv\hbar\omega V(\rho,\zeta)$,
the Eq.~(\ref{egp}) can be rewritten as
\begin{eqnarray}
i\frac{\partial\Phi}{\partial \tau}=\left[-\frac{1}{\rho}\frac{\partial}
{\partial\rho}\left(\rho\frac{\partial}{\partial\rho}\right)-\frac{\partial^2}
{\partial\zeta^2}+V(\rho,\zeta)-\mid\Phi\mid^{2}\right]\Phi ,
\label{egpa}
\end{eqnarray}
where the normalization of $\Phi$ is given by
\begin{equation}
\int_{-\infty}^\infty d\zeta \int_0^\infty d\rho \; \rho \mid \Phi\mid^2=
4\sqrt{2}\frac{N|a|}{l_0}\equiv 2 n .
\end{equation}
In the critical limit, we have $n=n_c=2\sqrt{2}k$. 

The equation, corresponding to Eq.~(\ref{stat}), for the 
chemical potential $\beta$ in dimensionless units
($\mu=\hbar\omega \beta$) is given by
\begin{equation}
\left[-\frac{1}{\rho}\frac{\partial}
{\partial\rho}\left(\rho\frac{\partial}{\partial\rho}\right)-
\frac{\partial^2}{\partial\zeta^2}
+V(\rho,\zeta)-|\Phi|^{2}\right] \Phi = \beta \Phi\,.
\label{tschd} 
\end{equation}     
As known \cite{Dalfovo}, the equation (\ref{tschd}) is valid in the mean-field 
approximation of the quantum many-body problem of a dilute gas, when 
the average inter particle distances are much larger than the absolute 
value of the scattering length and the wave-lengths are much 
larger than the average inter particle distance. 

The total energy of the system is given by
\begin{eqnarray}
E&=&\frac{N{\hbar\omega}}{2n}
\int_0^\infty \rho d\rho \int_{-\infty}^\infty d\zeta 
\left\{
\left[\frac{\partial}{\partial\rho} \Phi\right]^2 +
\left[\frac{\partial}{\partial z} \Phi\right]^2
\right. \nonumber \\ && \left.
+V(\rho,\zeta)\left| \Phi\right|^{2}
-\frac{|\Phi|^4}{2}
\right\} .  \label{Etot}
\end{eqnarray}

In order to analyze the effect of a deviation of the trap potential from 
the harmonic behavior, we consider two expressions for $V(\rho,\zeta)$,
labeled by $\nu=1, 2$ as follows:
\begin{equation}
V^{(\nu)}(\rho,\zeta)=
\frac{1}{4}\left[\omega_\rho^2\rho^2\left(1+\delta_\rho \rho^\nu\right)
+\omega_\zeta^2 \zeta^2\left(1+\delta_\zeta \zeta^\nu\right) \right] \,.
\label{potrap}
\end{equation}
The distortion added to the harmonic potential is cubic when $\nu=1$,
and quartic when $\nu=2$.
In both cases, the magnitudes
of the distortions in the directions $\rho$ and $\zeta$
are given by the parameters $\delta_\rho$ and $\delta_\zeta$.

\begin{figure}
 \setlength{\epsfxsize}{0.75\hsize} \centerline{\epsfbox{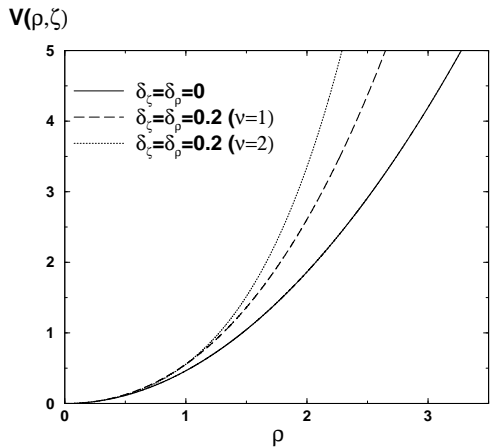}}
 \caption[dummy0]{ 
The anharmonic potentials, as given in Eq.~(\ref{potrap}), with
$\delta_\rho=0.2$ and $\zeta=0$, for the cubic ($\nu=1$, dashed-line) 
and quartic ($\nu=2$, dotted-line) case,
are compared with the harmonic potential ($\delta_\rho=0$, solid-line)
All the units are dimensionless, such that  
$U_{trap}=(\hbar\omega) V$ and 
$r_\perp=\rho\;l_{o}/\sqrt{2}$.
The values of the frequencies correspond to the ones given in 
Ref.~\cite{JILA}: $\omega_z=2\pi\times 6.8$ Hz and 
$\omega_{\perp}=2\pi\times 17.35 $ Hz.}
 \label{f0} 
 \end{figure}

With Eq. (\ref{potrap}), in both cases one can observe that
the interaction is approximately harmonic near the center of the trap
(at $\rho=0$ and/or $\zeta=0$). 
When $\rho<1$, the quartic term keeps approximately the harmonic 
shape of the potential in a more effective way than the cubic term.
In Fig. \ref{f0}, we can see how the shape of the harmonic trap 
($\delta_\rho=0$) changes in the radial direction, at the position 
$\zeta=0$, when one adds a cubic ($\nu=1$) or a quartic ($\nu=2$) term in 
the potential.

\section{Results obtained}

In this section, we report our main results considering the
solutions for the chemical potential, obtained for Eq.~(\ref{tschd}) with 
the anharmonic interaction given by Eq.~(\ref{potrap}), for several 
combinations of the parameters $\delta_\rho$ and 
$\delta_\zeta$, for $\nu=1$ and $\nu=2$. We have also considered
the solutions for the total energy, which is given
by Eq.~(\ref{Etot}).
For the numerical solution of Eq.~(\ref{tschd}), which is given in 
cylindrical symmetry, we have employed the Crank-Nicolson algorithm in 
two dimensions, employing the relaxation method propagating in the 
imaginary time, as earlier described in \cite{brief}. The spatial 
discretization used was 140 $\times$ 140, with cut-off 
$\rho_{max} = $ 7 and $\zeta_{max} = $ 7. The time discretization
was $\Delta\tau = $ 0.005 and the relaxation time used for obtaining the 
solutions of stationary ground state was $\tau = 16$. The results are more 
sensitive to the spatial grid spacing, but a lack of 
precision occurs in the third decimal figure of the values obtained in 
the numerical simulations. With these grids, we obtain precise 
results up to the second decimal figure, confirmed by the
convergence obtained with a refined spatial grid (160 $\times$ 160).   
We obtain our results for the total energy and the chemical potential of 
the condensate, as functions of the number of particles, by considering 
the geometry of the trap given in JILA's experiment reported in 
Ref.~\cite{JILA}, where
$\omega_z=2\pi\times $6.8 Hz and $\omega_\perp=2\pi\times $17.35 Hz. We 
consider a range of values for the parameters $\delta_\rho$ and $\delta_\zeta$,
that represents the corresponding anharmonicity. 
As shown, only positive anharmonic terms added to the original
harmonic trap can make the theoretical prediction for the maximum 
critical number $k$ become closer to the JILA's experimental result.
This implies that one should check for possible deviation of the
harmonic potential, outside the central region, that makes the trap more 
confining.

In Table I, we present our results for the maximum critical number
$k$, which is related to the critical number of atoms $N_c$, as 
given by Eq.~(\ref{1}), considering several possible choices for the
anharmonic parameters $\delta_\rho$ and $\delta_\zeta$, for both cases
$\nu=1$ (cubic) and $\nu=2$ (quartic). 

In the Fig. 2, we show some results for the case with $\nu=2$, where 
we have a quartic anharmonic term added to the original interaction,
in both directions $\rho$ and $\zeta$, such that 
$\delta_\rho=\delta_\zeta$. 
In this example, we are also considering the cigar-like symmetry 
of the trap used in Ref.~\cite{JILA}, with 
$\omega_\perp/\omega_z=17.35/6.8$.
We present two frames: for the total energy $E$ (lower frame), 
given in  units of $(N\hbar\omega)/(2n)$ and for the 
corresponding 
chemical potential $\mu$ (upper frame), in units of 
$\hbar\omega$, versus $N|a|/l_o$. 

This picture does not change too much when we switch off the deviation in the 
$\zeta$ direction, with $\delta_\zeta=0$ (the trap remains harmonic in the 
$z-$direction).
For the case with $\nu=1$, where we consider cubic anharmonic terms 
added to the original interaction, our results show that for 
both $\delta_z=\delta_\rho$ and $\delta_\zeta=0$ there are no significant 
qualitative differences in the observables plotted and the values are 
approximately equal to the quartic case. 
The behavior is similar for all the cases considered, with 
the collapse of the condensate occurring for smaller critical
number as we add larger positive deviation in the harmonic trap.
In this respect, the addition of a quartic term ($\nu=2$) in the
potential, is more effective in keeping the harmonic shape
near the center of the trap.

\begin{table}
\caption[dummy0]{Values of the maximum critical number $k$, for several 
combinations of the parameters $\delta_\rho$ and $\delta_\zeta$,
related to the anharmonic factors introduced in the interaction,
as given in Eq.~(\ref{potrap}), for $\nu=1$ (cubic) and
$\nu=2$ (quartic). The cylindrical symmetry is the same given in 
Ref.~\cite{JILA}.}
\begin{center}
\begin{tabular}{cccc}
$\nu$ & $\delta_\rho$ & $\delta_\zeta$ & $k$ \\ 
\hline
1& 0.0 & 0.0 & 0.550 \\
1& 0.1 & 0.1 & 0.529 \\
1& 0.1 & 0.0 & 0.532 \\
1& 0.2 & 0.2 & 0.512 \\
1& 0.5 & 0.5 & 0.476 \\
2& 0.1 & 0.1 & 0.520 \\
2& 0.1 & 0.0 & 0.525 \\
2& 0.2 & 0.2 & 0.497 \\
2& 0.5 & 0.5 & 0.456 \\
\end{tabular}
\end{center}
\end{table}

 \begin{figure}
 \setlength{\epsfxsize}{1.0\hsize} \centerline{\epsfbox{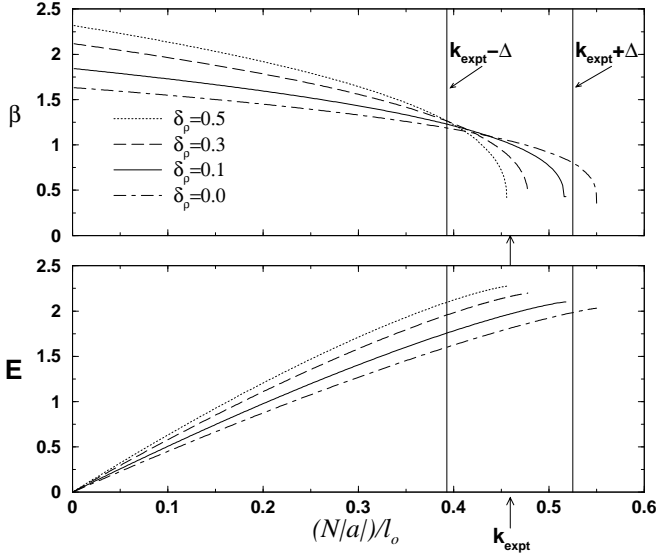}}
 \caption[dummy0]
{Total energies $E$ (lower frame), in units of $(N\hbar\omega)/(2n)$ 
[$E_{total}\equiv E (N\hbar\omega)/(2n)$], and chemical potentials 
$\mu$ (upper frame), in units of $\hbar\omega$ [$\mu\equiv 
\beta(\hbar\omega)$], versus $N|a|/l_o$, for different values
of $\delta_\rho=\delta_\zeta$, as given inside the upper frame.
The deviations from the harmonic trap are given by quartic terms in both 
directions $\rho$ and $\zeta$. 
Here, we are also using the trap considered in Ref.~\cite{JILA}, with
the asymmetry given by $\omega_\perp/\omega_z=2.5515$.
The corresponding critical numbers are at the end of each curve,
with the position of the experimental value $k_{expt}$ pointed out with
vertical arrows. $\Delta$ corresponds to the sum of the experimental
errors. }
 \label{f4}
 \end{figure}

 \begin{figure}
 \setlength{\epsfxsize}{1.0\hsize}\centerline{\epsfbox{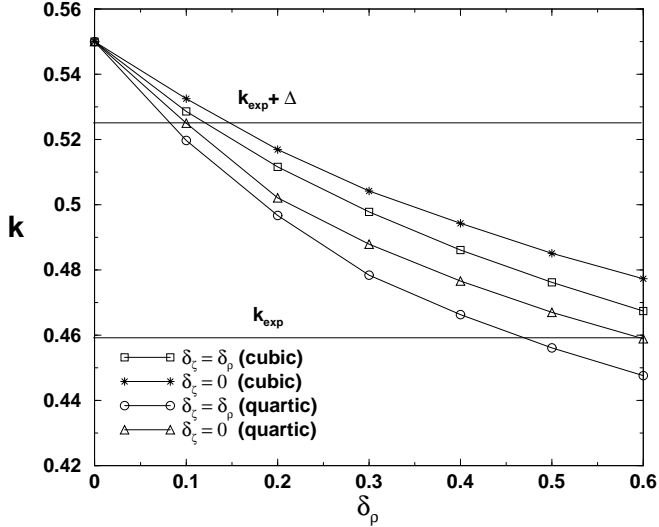}}
 \caption[dummy0]{Variations of the critical number of atoms, $k$, in the 
condensate, as functions of $\delta_\rho$.
 The corresponding deviations in the $\zeta$ direction, $\delta_\zeta$, 
are given inside the figure. The position of the experimental $k$ and the 
corresponding positive error bar ($\Delta$) are also indicated.}
 \label{fc}
 \end{figure}

All the above results can be globally analized by examining Fig. 3, in      
which we have the behavior of the maximum critical number of atoms in the   
condensate state, parametrized by $k$, as a function of the magnitude of the 
trapping anharmonic potential, parametrized by $\delta_\rho$. We display the 
effect of both cubic and quartic anharmonicity in confinement, with or 
without including the deviation in the z-direction.
As we can see, the higher is the anharmonic parameter, the smaller is the 
critical number of condensed atoms. As one can observe, in order to obtain 
the theoretical results for $k$ inside the region covered by the experimental 
error bars, one needs deviations of about 10\% to 20\% from the harmonic 
trap, at distances of the order of the oscillator length.

\section{Conclusions}
In summary, we have solved numerically the stationary Gross-Pitaevskii 
equation in cylindrical symmetry, for condensed systems with attractive 
two-body interaction, with anharmonic trapping potentials, either with 
cubic and quartic deviation from the harmonic oscillator. 
We did the present study in the perspective to observe the real 
effect on $N_c$ of anharmonicities that can occur in the usual 
confining traps that have been extensively employed in the experiments
with atomic BEC. We took the harmonic trap parameters considered in 
Ref.~\cite{JILA}, to exemplify how much should be the strength of 
cubic or quartic anharmonic terms in the trap interaction to present a 
sizable change in the critical parameter $k$; in the perspective of
a partial explanation of the observed experimental deviations from the
previous theoretical estimates. 
The present results show that only with a deviation of about 10\% from the 
harmonic potential, at distances of the order of the oscillator length,
it becomes possible to explain theoretically the experimental results 
reported in \cite{JILA}, within the error bars. 
However, this is quite large strength for the anharmonic term, that is not 
supported by the detailed analysis and description of the experiments 
given in Ref.~\cite{Roberts}. 

In a more realistic perspective, our present work is predicting the 
quantitative effect in the critical number of atoms due to 
anharmonicities that can occur in a trap. For the non-harmonic terms
we have considered cubic or quartic ones. In cases of other shapes
for the anharmonic terms, one can also make qualitative predictions
based on the present results. We should point out that our study is
not restricted to the example (trap frequencies and symmetry) that 
we have used; it can be easily extended to other trap arrangements, 
and can be useful in experimental analysis of condensates with attractive 
two-body interactions. In this respect, we  should note that it has been 
studied and reported the possible occurrence 
of anharmonicities in time orbiting potential traps~\cite{minogin}.
Our work is considering realistic situations, that are close to the
actual experiments.
So, our aim in the present work, was to report the effect of a deviation 
in the harmonic trap that increases as we go to regions outside the center 
of the system. 
    
We would like to thank Prof. Rajat K. Bhaduri for useful discussions 
and suggestions, and Dr. Elizabeth Donley for pointing us the Ref. 
\cite{Roberts}.  
This work was partially supported by Funda\c c\~ao de Amparo \`a Pesquisa 
do Estado de S\~ao Paulo (FAPESP). LT also thanks support from Conselho
Nacional de Desenvolvimento Cient\'\i fico e Tecnol\'ogico (CNPq) 
of Brazil.

\end{document}